\documentclass[a4paper,10pt]{iopart}

\usepackage[utf8]{inputenc} 
\usepackage{graphicx}  

\expandafter\let\csname equation*\endcsname\relax
\expandafter\let\csname endequation*\endcsname\relax
\usepackage{amsmath}

\usepackage{dsfont}
\usepackage{textcomp}
\usepackage[normalem]{ulem} 
\usepackage{float}
\usepackage{units}
\usepackage{listings}
\usepackage[format=hang, font={footnotesize, sf}, labelfont={bf}, margin=1cm, aboveskip=5pt, position=bottom]{caption}
\usepackage{placeins}
\usepackage{cite}

\newcommand{\ud}{\,\mathrm{d}}
\newcommand{\vek}[1]{\vec{#1}}

\newcommand{\nabpara}{\ensuremath{\nabla_{\parallel}}} 
\newcommand{\laplace}{\ensuremath{\nabla^2}}

\newcommand{\neut}[1]{#1}
\newcommand{\curv}[1]{#1}
\newcommand{\diff}[1]{#1}
\newcommand{\conv}[1]{#1}
\newcommand{\esta}[1]{#1}
\newcommand{\tpho}[1]{#1}
\newcommand{\fric}[1]{#1}
\newcommand{\pres}[1]{#1}

\newcommand{\dens}[1]{$n_0=#1\cdot10^{18}$\,m$^{-3}$}
\newcommand{\MW}[1]{#1\,MW/m$^2$}

\graphicspath{{images/}{}}

\providecommand{\bibAnnoteFile}[1]{}
\providecommand{\bibAnnote}[2]{}

\newcommand{\TITLE}{Dynamics of scrape-off layer filaments in detached conditions}
\newcommand{\TITLEL}{\TITLE} 

\providecommand{\AUTHORL}{David Schworer}

\usepackage[pdftex,%
	    pdftitle=\TITLEL,%
	    pdfauthor=\AUTHORL,%
	    colorlinks,%
	    citecolor=black,%
	    filecolor=black,%
	    urlcolor=black,%
	    linkcolor=black]{hyperref}
\begin{document}

\title{\TITLE}

\author{  D.~Schw{\"o}rer${}^{a,b}$, N.~R.~Walkden${}^{b}$, B.~D.~Dudson${}^{c}$,
  F.~Militello${}^{b}$, H.~Leggate${}^{a}$, 
  M.~M.~Turner${}^{a}$}

\address{
      ${}^{a}$Dublin City University, Dublin 9, Ireland\\
      ${}^{b}$CCFE, Culham Science Centre, Abingdon, 
      Oxfordshire, OX14 3DB, UK\\
      ${}^{c}$ York Plasma Institute, Department of Physics,
University of York, Heslington, York YO10 5DQ, UK}
\ead{david.schworer2@mail.dcu.ie}
\vspace{10pt}
\begin{indented}
\item[]\today
\end{indented}



\begin{abstract}
The here presented work studies the dynamics of filaments using 3D
fluid simulations in the presence of detached background
profiles. It was found that evolving the neutrals on the time-scale of
the filament did not have a significant impact on the dynamics of the filament.
In general a decreasing filament velocity with increasing plasma background
density has been observed, with the exception of detachment onset,
where a temporarily increase in radial velocity occurs. The decreasing trend with temporary increase was
found for filaments around the critical size and larger, while smaller
filaments where less affected by detachment.
With detachment the critical filament size increased, as larger filaments were
faster in detached conditions. This breaks the trend of attached conditions,
where the critical size decreases with increasing density.
\end{abstract}

\section{Introduction}
Filaments are field aligned pressure perturbations, observed in the
tokamaks scrape-off layer (SOL), that have a much
higher amplitude than the background fluctuations. They have been
observed in most magnetized plasmas, including most fusion devices,
where various properties of SOL fluctuations have been measured, for
example the skewed
probability distribution
function~\cite{dippolito11a,walkden17a,sanchez00a,antar01a,graves05a,garcia07a,boedo09a,kube18a}.
In the SOL of tokamaks, filaments have been observed to cause
a significant amount of transport across the magnetic
field~\cite{boedo01a,carralero18a,kube19a}.
Filaments can be modelled non-linearly in two dimensions, evolving two fields,
namely the density and the
vorticity~\cite{blob2dpy19a,dippolito11a,easy14a}. In this case the
lack of
the third dimension, parallel to the magnetic field, is typically
closed by the advection closure or the sheath
closure~\cite{easy14a,krasheninnikov01a,garcia06a,theiler09a,kube11a}. This
has
been extended in various ways in recent years, e.g. including the full
parallel
dynamics, which is required for capturing drift waves, and capturing
sheath dynamics~\cite{easy14a,angus12a,jovanovic08a}.

The dynamics of filaments has a dependence on their perpendicular size
$\delta_\perp$. While for small filaments a significant amount of the
diamagnetic current across the filament is closed via currents in the
drift-plane, sheath currents
are important for large filaments~\cite{dippolito11a,easy14a}.  The
radial velocity has a maximum at a specific size - referred to as the
critical size $\delta^*$, which represents a balance point between
these two vorticity sinks.

A major challenge for the operation of fusion devices, such as ITER,
is the power handling in the divertor. ITER will thus have to operate
in detached, or at least partially detached conditions~\cite{potzel14a}.
Detachment is an operational regime in which the heat and particle target fluxes are
reduced, as a significant part of the plasma is cooled before it can
reach the target~\cite{hutchinson94a,wischmeier09a}.
Detached conditions require a drop of total plasma pressure along the
flux tube.
Charge-exchange can be an efficient sink for plasma momentum as well
as a sink for the plasma energy, reducing the heat
load at the target, as well as reducing the density flux.
A common condition to define detachment is the so called
roll-over~\cite{coster11a,potzel14a}. As the upstream density is
increased, during attached operation the particle target flux
increases. As detachment is reached, the particle target flux drops
with increasing upstream density, as a result of the loss of pressure
in the vicinity of the target due to the strong interaction with the
neutrals.

As detachment is reached the plasma temperature drops significantly in
the vicinity of the target. This causes an increased plasma
resistivity. It has been predicted that an increased resistivity
results in an increased radial filament velocity~\cite{krasheninnikov08a,easy16a,garcia09a}.
Easy\etal\cite{easy16a} found by introducing an
artificially increased resistivity 
in addition to the increased radial velocity also an increased critical
size $\delta^*$ with increasing resistivity~\cite{easy16a}.
The here presented study extends the study by Easy\etal\cite{easy16a}
in a self consistent way.

Filaments are a significant cross field transport mechanism and have
been shown that they significantly influence the time averaged
profiles~\cite{militello16c,militello18a}. At the same time filaments
depend on the background condition~\cite{schworer18a}.
Our previous study found the importance of realistic background
profiles for the dynamics of filaments, where a strong dependence of
the target temperature has been observed in attached
conditions~\cite{schworer18a}.
In the attached conditions, only a weak influence of
neutrals on the dynamic of filaments has been observed.
The increased resistivity by a decreased temperature did not result in
an increased velocity, as the change in target temperature had a
stronger effect~\cite{schworer18a,easy16a}

This work aims to extend this into detached conditions, as it is
expected that the filament-neutrals interaction will become important
in the higher density cases.
This also allows to extend the resistivity of the background plasma in
a self-consistent way to levels studied by Easy\etal\cite{easy16a}.

The model used will be introduced in sec.~\ref{s:model}.
After a short introduction to the backgrounds profiles in
sec.~\ref{s:bg}, the results of the simulations are presented in
sec.~\ref{s:evo}.
The simulation include the direct influence of neutrals
(sec.~\ref{s:di}), the impact of detachment on the radial velocity
(sec.~\ref{s:de}) and on the critical size (sec.~\ref{s:ds}) as well as
the rigidness of filaments (sec.~\ref{s:r}).
The results will be discussed in sec.~\ref{s:dis}, before a short
summary is given ins sec.~\ref{s:sum}.


\section{Model}\label{s:model}
STORM is a 2 fluid plasma model for the study of
filaments~\cite{easy14a,easy16a,walkden16a,schworer17a} implemented
using the MPI parallelised library BOUT++~\cite{dudson08a,dudson15a}.
BOUT++ allows a user to implement fluid models in curvi-linear geometry in
close to analytical form. BOUT++ is open source, published under the
LGPL license, and thus freely
available~\footnote{https://boutproject.github.io/}.
STORM has recently been extended to include
neutrals~\cite{schworer17a,schworer18a}.

Similar to the previous study, 1D background profiles, an extension of
the two-point model, only retaining the parallel dimension are evolve
to steady state. The 1D domain represents a flux tube, with one end
representing upstream, and the other one representing the target with
sheath boundary conditions. Onto the profiles the filaments are seeded
and evolved.

The neutral model consists of the density of the neutrals atoms
$n_n$, of which the logarithm is evolved. The equations are written in
Bohm units, as is the rest of the STORM code~\cite{schworer18a,easy14a}.
\begin{align}
 \frac{\partial log(n_n)}{\partial t}  &= \frac{1}{n_n} \big(
  - \nabla_\parallel m_n + S_R
  + \Gamma^\text{rec} - \Gamma^\text{ion}
  + \nabla_\parallel \mu_{n_n} \cdot \nabla_\parallel n_n \big) - f_l
   \intertext{and the velocity $v_n$ along the magnetic field lines, for
 which the momentum $m_n$ is evolved}
 \frac{\partial m_n}{\partial t} &= - \nabla_\parallel v_n m_n
    - \nabla_\parallel n_n T_n + \Gamma^\text{rec} U \nonumber\\
  &\quad
    + \Gamma^\text{CX} (U - v_n)  - \Gamma^\text{ion} v_n +
    \mu_{m_n} \nabla_\parallel^2 m_n - S_R v_{th} \big)
\end{align}
The charge exchange, ionisation and recombination rates are
denoted by $\Gamma^\text{CX}$, $\Gamma^\text{ion}$ and
$\Gamma^\text{rec}$ respectively.
\begin{figure}
  \centering
  \includegraphics[width=.7\linewidth]{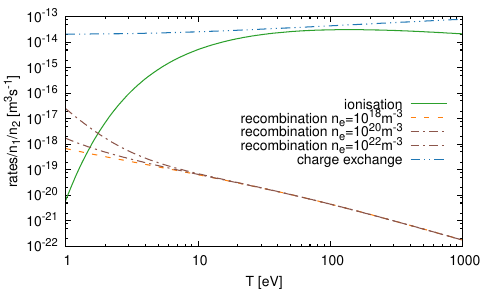}
  \caption{Temperature dependence of the neutral interaction rates
    $\Gamma^\text{CX}$, $\Gamma^\text{ion}$ and
  $\Gamma^\text{rec}$.}\label{f:rates}
\end{figure}
The temperature dependence is shown in fig.~\ref{f:rates}.
$U$ denotes the ion parallel velocity.
The model is similar to the UEDGE fluid model~\cite{wising96a}.
As the plasma flows are strongly field aligned, evolving the parallel
velocity of the neutrals allows to model the friction in a momentum
conserving way. As the background are only one dimensional, the
inclusion of perpendicular velocities of the neutrals would thus have
no impact on the backgrounds.
$S_R(x,y,z_\parallel,t) \propto exp((z_\parallel-z_t)^2/l_R^2)\cdot
n_tU_t$ is the source due to recycling, which is a Gaussian function
located at the target $z_\parallel=z_T$ with falloff length $l_R$ of 0.1\,m.
The integral $\int S_R \ud z=0.99 n_tU_t$ sums to 99\,\% of the
particle target flux $n_tU_t$. $\mu_{n_n}$ is the neutral diffusion, and $f_l$ is
a loss fraction, compensating the lack of cross-field losses.
The temperature of the neutrals $T_n$ is not evolved, and is assumed
to be 3\,eV, close to the Franck-Condon energy~\cite{dudson17a}, with the
associated thermal speed $v_{th}$.
Evolving the logarithm of the density, rather then the density itself
is beneficial for the stability of the code. As the neutral
density can vary quite strongly within a small spatial region, this is
of importance for running simulation in detached conditions. The
logarithm further ensures that the density-solution always remains
positive. Using the logarithm also changes the error norm, which
matters as an iterative solver (PVODE) is used to evolve the
system. Using the logarithm
increases the accuracy in the presence of low
densities.
For the recycling, in previous models an exponential function
$S_R \propto exp((z_\parallel-z_t)/l_R)$ was
chosen. As a small recycling falloff length $l_R$ is preferred to model the recycling in
detached conditions, this caused a strong finite-size dependence, as
the gradient is highest next to the target, thus refining the grid
causes the neutrals to be deposited increasingly close to the the
target. This is avoided by switching to a Gaussian function. It still
retains the strong drop-off further away from the target.
As later shown in fig.~\ref{f:bg}, the strong gradients near the
sheath boundary conditions, due
to detachment, need to be resolved. Thus the mesh contained 480
points in parallel direction, giving a uniform grid spacing of
$\approx 2.1$\,cm, significantly refined compared to previous
studies~\cite{schworer18a,easy14a}.

The geometry used is a simple slab geometry, with $x$ being the radial
coordinate, $z$ being the parallel direction, and $y$ being the
bi-normal direction. Only half of the flux tube is simulated, thus
symmetry boundary conditions are used at the
mid-plane~\cite{schworer18a}.  At the target sheath boundary
conditions are applied, requiring the ion velocity to reach the speed
of sound $c_s$, and the electrons $c_s \exp(-V_f-\frac \phi T)$ with
the floating potential $V_f = \frac 1 2 log\left(\frac{2 \pi}{\mu}\right)$
and the ion-electron mass ratio is $\mu=m_i/m_e$.

The STORM equation describing the plasma consist of the density $n$
equation for the electrons
\begin{align}
     \frac{\partial n}{\partial t} &=
          \conv{\frac{\nabla \phi \times \vek b}{B} \cdot \nabla n}
          - \conv{\nabla_\parallel (V n)}
          + \diff{\mu_n \laplace n}
          - \curv{g n \frac{\partial \phi}{\partial y}}
          + \curv{g \frac{\partial n T}{\partial y}}
          + \neut{\Gamma^\text{ion}}
          - \neut{\Gamma^\text{rec}}
          \\[0.5ex]
          \intertext{where $\phi$ denotes the potential, which is the
          Laplacian inversion $\omega=\laplace_\perp \phi$ of the
          vorticity.
          The magnetic field is of direction $\vec b=\hat z$ and of magnitude
          $B=0.5$\,T.
          $\mu_\alpha$ is the diffusion rate for quantity $\alpha$, so
          $\mu_n$ denotes the diffusion rate for the density.
          $g$ is the effective gravity constant playing the role of
          magnetic curvature. $g$ is related to the major
          radius $R_c$ which is set to $1.5$\,m thus
          $g=\frac{2}{R_c}\approx 1.33$\,m$^{-1}$.
          Both $R_c$ and $B$ were chosen to be representative of MAST.
          The terms
          containing $g$ are drive terms.
          The equation for the parallel electron
          velocity $V$ is given by}
     \frac{\partial V}{\partial t} &=
          \conv{\frac{\nabla \phi \times \vek b}{B} \cdot \nabla V}
          - \conv{V \nabpara V}
          + \esta{\mu \nabpara \phi}
          - \pres{\frac{\mu}{n}\nabpara nT}
          \\
          &\quad
          + \fric{n \mu \eta_\parallel (U-V)}
          - \tpho{0.71 \mu\nabpara T}
          - \neut{\frac V n \Gamma^\text{ion}}
          + \mu_\parallel \nabpara^2 V \nonumber
          \\
          \intertext{The parallel ion-electron resistivity is
            given by $\eta_\parallel$.
            Electron neutral collisions are neglected, as they only
            become important below 1\,eV~\cite{easy16a}.
            The viscosity term
          $\mu_\parallel$ was introduced to improve the numerical
          stability, with the magnitude
          $\hat\mu_\parallel=1600 m^2s^{-1}$ or in Bohm units
          $\mu_\parallel=20$ well below the Braginskii
          level~\cite{wesson04a}.
          The equation for the parallel ion
          velocity $U$ is given by}
     \frac{\partial U}{\partial t} &=
          \conv{\frac{\nabla \phi \times \vek b}{B} \cdot \nabla U}
          - \conv{U \nabpara U}
          - \esta{\nabpara \phi}
          - \fric{\eta_\parallel n (U-V)}
          \\
          &\quad
          + \tpho{0.71 \nabpara T}
          - \neut{\frac{U} n \Gamma^\text{ion}}
          - \neut{\frac U n \Gamma^\text{CX}}
          + \mu_\parallel \nabpara^2 U\nonumber
          \\
          \intertext{the equation for the electron temperature $T$
            is given by}
     \frac{\partial T}{\partial t} &=
          \conv{\frac{\nabla \phi \times \vek b}{B} \cdot \nabla T}
          - \conv{V \nabpara T}
          + \frac{2}{3} \left( \frac{-1}{n}\nabpara{q_\parallel}
            + \tpho{0.71 (U-V) \nabpara T}
            - T \nabpara V\right. \label{eq:temp}
          \\
          &\quad\left.
            + \diff{\frac{\kappa_\perp}{n} \laplace_\perp T}
            + \fric{\eta_\parallel n (U-V)^2}
            + \mu_\parallel \frac V n \nabpara^2 V
            \right )
          - \curv{\frac 2 3 g T \frac{\partial \phi}{\partial y}}
          - \curv{\frac 2 3 g \frac{T^2}{n} \frac{\partial n}{\partial y}}\nonumber
          - \curv{\frac 7 3 g T \frac{\partial T}{\partial y}}
          \\
          &\quad
          - \curv{\frac 2 3 g V^2 \frac{1}{\mu n} \frac{\partial n T}{\partial y}}
          - \neut{\frac{T}{n} \Gamma^\text{ion}}
          - \neut{\frac{T}{n} \Gamma^\text{CX}}
          - \neut{\frac{1.09 T-13.6 eV}{n} \Gamma^\text{rec}}
          - \neut{\frac{30 eV}{n} \Gamma^\text{ion}}
          - \neut{R^\text{imp}}
          \nonumber
          \intertext{The parallel heat conduction is
          given by $q_\parallel$ and $\kappa_\perp$ is the
          perpendicular heat transport coefficient.
          The impurity radiation $R^\text{imp}$ is using the carbon
          radiation model from Hutchinson~\cite{hutchinson94a} using
          an impurity fraction of 1\,\%.
          The radiation seems to have only a minor impact, and is not
          responsible for the total plasma pressure drop in front of
          the target. 
          The equation for the vorticity $\omega$ is given by}
          \frac{\partial \omega}{\partial t} &= A_\omega +
          J_\omega + B_\omega + C_\omega + N_\omega
 \label{eq:vort}\\
 A_\omega &= \conv{\frac{\nabla \phi \times \vec b}{B} \cdot
   \nabla \omega} - \conv{U \nabla_\parallel \omega}\\
 J_\omega &= \esta{\nabla_\parallel (U-V)}
          + \esta{\frac{U-V}{n}\nabla_\parallel n}\\
 B_\omega &= \diff{\mu_\omega \laplace \omega}
          + \diff{\nabla_\perp \mu_\omega \cdot \nabla_\perp \omega}\label{eq:vort:B}\\
 C_\omega &= \curv{\frac g n \frac{\partial n T}{\partial y}}\\
 N_\omega &= - \neut{\frac{1}{n}\laplace_\perp \phi (\Gamma^\text{CX} + \Gamma^\text{ion})}
          - \neut{\frac{1}{n}\nabla_\perp \phi \cdot \nabla_\perp(\Gamma^\text{CX} + \Gamma^\text{ion})}\label{eq:vort:N}
          \intertext{with $A_\omega$ the advective contribution of
            the vorticity equation, $J_\omega$ the one arising from
            the current, $B_\omega$ the diffusive and $N_\omega$ the
            contribution due to neutral plasma interaction.  The
            curvature drive terms are $C_\omega$. $\omega$ gives the
            potential $\phi$ using the Boussinesq approximation}
          \omega &= -\laplace_\perp \phi
          \intertext{with the viscosity
            $\mu_\omega$ given by }
          \mu_\omega &=
          (1+1.6 q^2)\frac 6 8
          \frac{\rho_i^2 n  Z^4\Lambda}{\sqrt{m_i}\epsilon^2_0 3(2\pi
            T_i)^{1.5}}\propto \frac{n}{T^{\frac 1 2}}\label{eq:mu}
\end{align}
The cross field transport coefficients are calculated self consistently,
following the derivation of
Fundamenski~\etal~\cite{fundamenski07a,schworer18a}.
Further details about the model are given in
reference~\cite{schworer18a}.
The neutral terms in the vorticity equation~\eqref{eq:vort} provide a
sink for the drive. In order to derive a scaling for this sink,
we follow the common approach and balance the drive term with the neutrals
sink~\cite{krasheninnikov01a,myra06b,theiler09a,kube11a,kube12a,manz13a,walkden16a}:
\begin{align}
  \frac g n \frac{\partial n T}{\partial y} &\sim \neut{\frac 1 n \laplace_\perp
    \phi (\Gamma^\text{CX} + \Gamma^\text{ion})}
  +\neut{\frac 1 n \nabla_\perp \phi \cdot
    \nabla_\perp(\Gamma^\text{CX} + \Gamma^\text{ion})}
  \intertext{Replacing derivatives by the inverse filament size}
  g \frac{\delta_p}{\delta_\perp} & \sim
  \frac{\phi \left(\Gamma^\text{CX} +
    \Gamma^\text{ion}\right)}{\delta_\perp^2}
  \intertext{with the pressure perturbation $\delta_p$}
  \phi &\sim \delta_p \frac{g\delta_\perp}{\Gamma^\text{CX} +
    \Gamma^\text{ion}}
  \intertext{we get the filament velocity}
  v_r &\sim \delta_p \frac{g}{\Gamma^\text{CX} +
    \Gamma^\text{ion}}
\end{align}
Thus if the curvature drive is balanced mainly by the neutrals, the filament
velocity should have no size dependence $v_r \propto \delta_\perp^0$.
For filaments in the viscous regime, where the drive is balanced by
viscosity, a scaling of $v_r \propto \delta_\perp^2$ is
expected~\cite{easy16b}.
In the inertial regime, a scaling of $v_r \propto \delta_\perp^\frac 1 2$
is expected, and in the sheath limited regime, a scaling of
$v_r \propto \delta_\perp^{-2}$ is expected, thus the neutral scaling lies between
the inertial and sheath limited regime, and is expected to be the
dominant closing mechanism for large filaments, that are not able to
connect to the sheath.

\begin{figure}
  \centering
  \includegraphics[width=\linewidth]{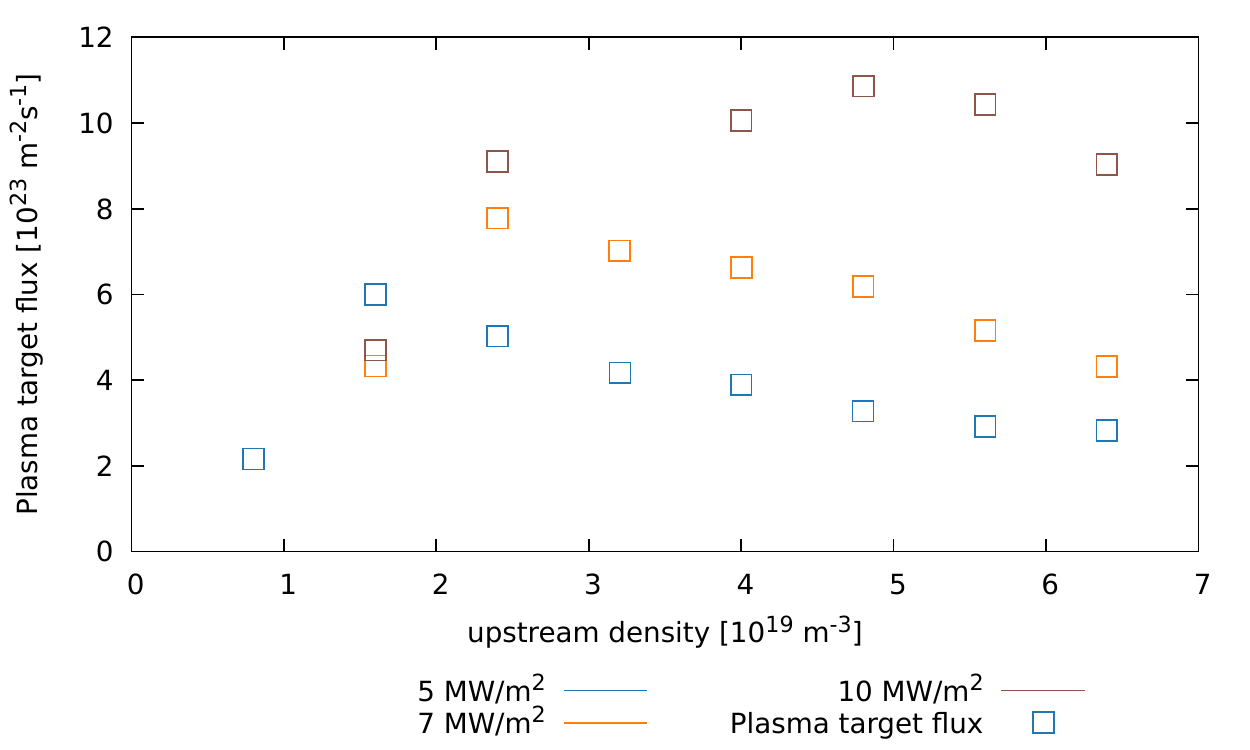}
  \caption{The first three base vectors, representing monopole,
    radial bipole and bi-normal dipole.}\label{f:ab}
\end{figure}

The different contribution of the vorticity can be ether stored
during the simulation, or calculated as part of the post-processing.
The new python interface for BOUT++ allows to use exactly the same
numerical implementation, that was used during the
simulation in the post-processing scripts~\cite{BOUTv4-3-0}.
This results in several 4D data sets that have to be compared. By
restricting to the analysis to $t=t_\text{peak}$ where the filaments
are fastest, restricts the data set to 3D. The structure of the
contribution in the perpendicular plane is of
interest~\cite{walkden16a}.
In order to further reduce the the amount of data, each 2D slice $d$
in the perpendicular plane was projected onto a set of physically
motivated basis vectors $e_{i,j}$.
The basis vectors $e_{i,j}$ are centred on the centre of mass of the filament,
and periodicity in $y$ was taken into account such that the
branch-cut was at $y_\text{cut}=c_y\pm \frac{1}{2}L_y$.
The first 3 base vectors
\begin{align}
  e_{0,0}&=\sqrt\frac{2}{\pi}\delta_\perp^{-1}
  e^{-\frac{x^2+y^2}{\delta_\perp^2}}\\
  e_{1,0}&=\sqrt\frac{2}{\pi}\frac{2x}{\delta_\perp}\delta_\perp^{-1}
  e^{-\frac{x^2+y^2}{\delta_\perp^2}}\\
  e_{0,1}&=\sqrt\frac{2}{\pi}\frac{2y}{\delta_\perp}\delta_\perp^{-1}
  e^{-\frac{x^2+y^2}{\delta_\perp^2}}
\end{align}
are shown in fig.~\ref{f:ab}

\begin{figure}
  \centering
  \includegraphics[width=\linewidth]{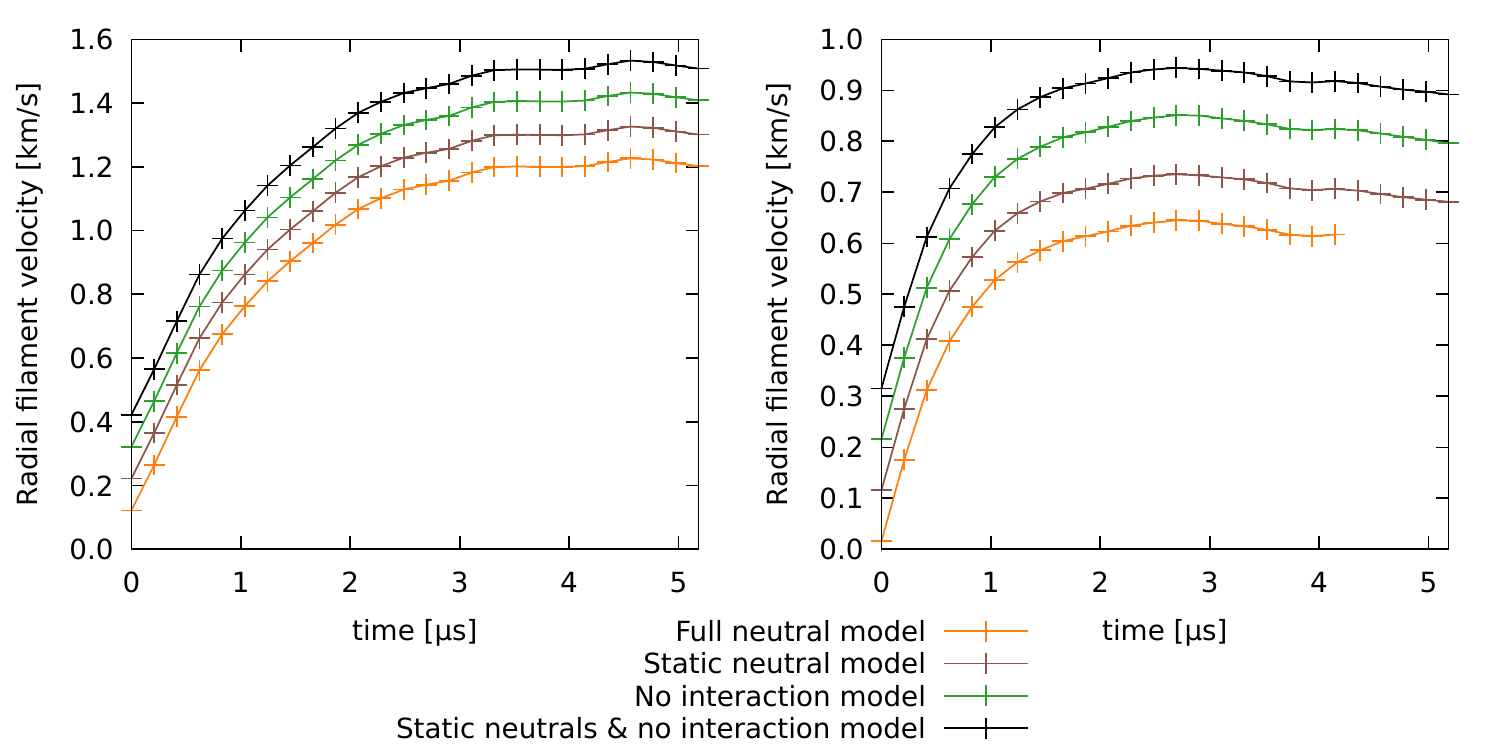}
  \caption{Some 2D simulation data and its projection on the three
    base vectors shown in fig.~\ref{f:ab}.
    Rather then the projection itself, the projection multiplied
    by the basis vector is shown.
    All quantities are
    plotted on the same scale. It can be seen that the monopole
    contribution is rather weak, with $-3.0 e_{0,0}$ a.u. The
    radial contribution is larger with $-5.9 e_{1,0}$ a.u. while the
    bi-normal contribution is the largest one with $18 e_{0,1}$ a.u.
  }\label{f:ad}
\end{figure}

The data $d$ can be projected onto the basis vectors, thus we get the
data components $d_{i,j}$ for each slice in $z$:
\begin{align}
  d_{i,j}(z) &= e_{i,j} \cdot d  = \int\int e_{i,j}(x,y) d(x,y,z)
  \ud x \ud y
\end{align}
Probably the most interesting component is the (0,1) component which
represents a dipole in bi-normal direction. This creates an $E\times
B$ force in the radial direction, and is thus related to the radial
drive.
The other two moments that have been analysed are the (0,0) component,
that represents the monopole contribution, and the (1,0) component,
that in turn describes a dipole in radial direction.
A monopole in the vorticity field is causing a monopole potential, and
thus via $E\times B$ a spinning motion~\cite{walkden16a}.
The (1,0) component, or dipole in radial direction causes a motion in
bi-normal direction.
Fig.~\ref{f:ad} shows the decomposition of some data, that would
represent a single point in the later discussed fig.~\ref{f:vc1} in
each of the three plots for monopole, bi-normal dipole and radial
dipole.



After the backgrounds have been calculated, the neutral-plasma
interaction was modified in the filament simulations, to study the
filament-neutrals interaction in more detail. In the ``full neutral''
model, the neutrals are evolved, using the equations above, and the
interaction rates $\Gamma^\alpha$ are calculated self
consistently. In the ``static neutral'' model, the neutrals are kept at their
steady state values as determined by the background calculations, and
the interaction rate $\Gamma^\alpha$ are
calculated based on plasma and neutral distribution. In the ``no
interaction'' model, the neutral contribution to the vorticity
equation is dropped.

\section{Background profiles}\label{s:bg}
In order to run the filament simulations, 1D backgrounds along the
parallel direction were computed using the equations described in
section~\ref{s:model}.  An upstream energy source is included which is
exponentially shaped and located at the mid-plane, as well as a
Gaussian density source, which is also located at the mid-plane.  The
density source was controlled in such a way that a
predefined density value upstream was achieved.
The energy influx is set to values in the range of \MW{5} to \MW{10} and
\begin{figure}
  \centering
  \includegraphics[width=\linewidth,page=1]{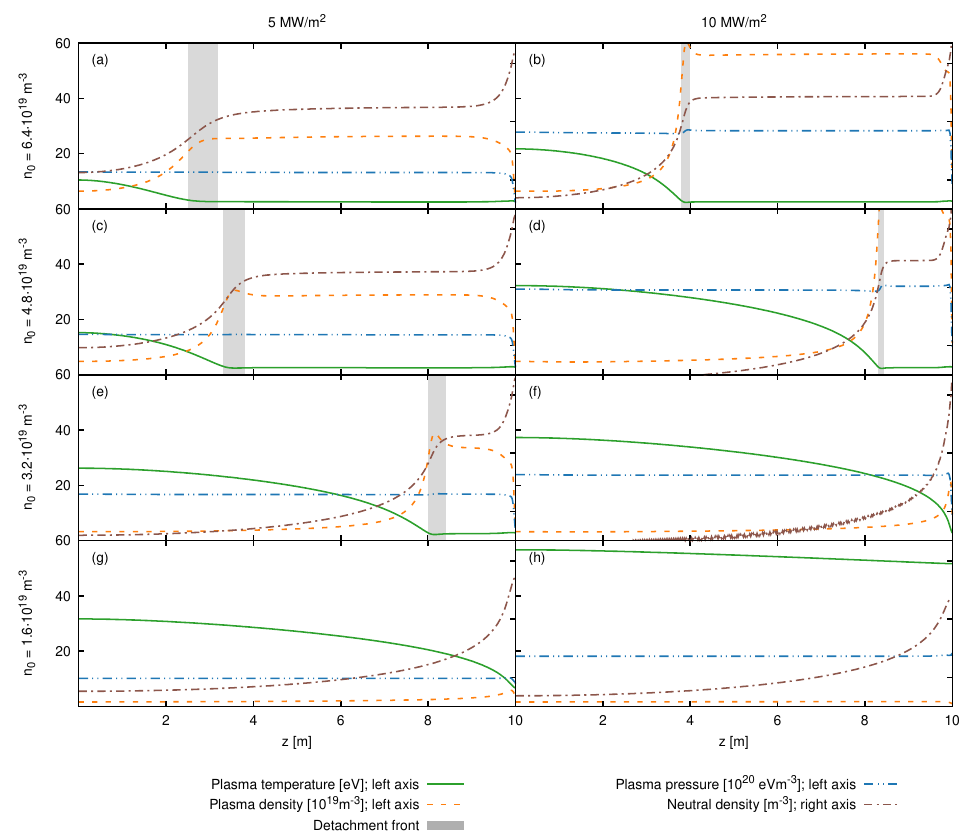}
  \caption{
    Background plasma profiles, run to steady-state for a set
    energy influx $P_i$ and density $n_0$. The sheath is at the right hand
    side at $L=10$\,m. The mid-plane is at the left side, and is a
    symmetry plane.
    The scales are kept the same in all sub plots, to allow an easy
    comparison of the differences.
    Plasma
    density $n_p$ (green line) in units of $10^{19}$\,m$^{-3}$ and
    temperature $T$ (orange dashed) in units of eV
    is plotted to the linear scale on the left hand side.
    The neutral density $n_n$ (brown dash dotted) is plotted to
    the log scale on the right hand side.
    Plasma pressure $p_n$ (blue dash double dotted) in units of
    $2\cdot10^{19}$\,eVm$^{-3}$ is the sum of static plasma pressure
    $n_pT$ and dynamic pressure $nU^2$, plotted on the linear scale on
    the left.
    On the left hand side of the plot are simulations with an energy
    influx of 5\,W/m$^2$, right
    are profiles with an energy influx of 10\,W/m$^2$.
    The upstream densities $n_0$ are bottom to top
    $n_0=16\cdot10^{18}, 32\cdot10^{18}, 48\cdot10^{18}$ and
    $64\cdot10^{18}$\,m${-3}$.
    The detachment front is highlighted by a shaded area.
    Fig.~\ref{f:bg2} provides a zoom-in of the last m in front of
    the target.
    }\label{f:bg}
\end{figure}
\begin{figure}
  \centering
  \includegraphics[width=\linewidth,page=2]{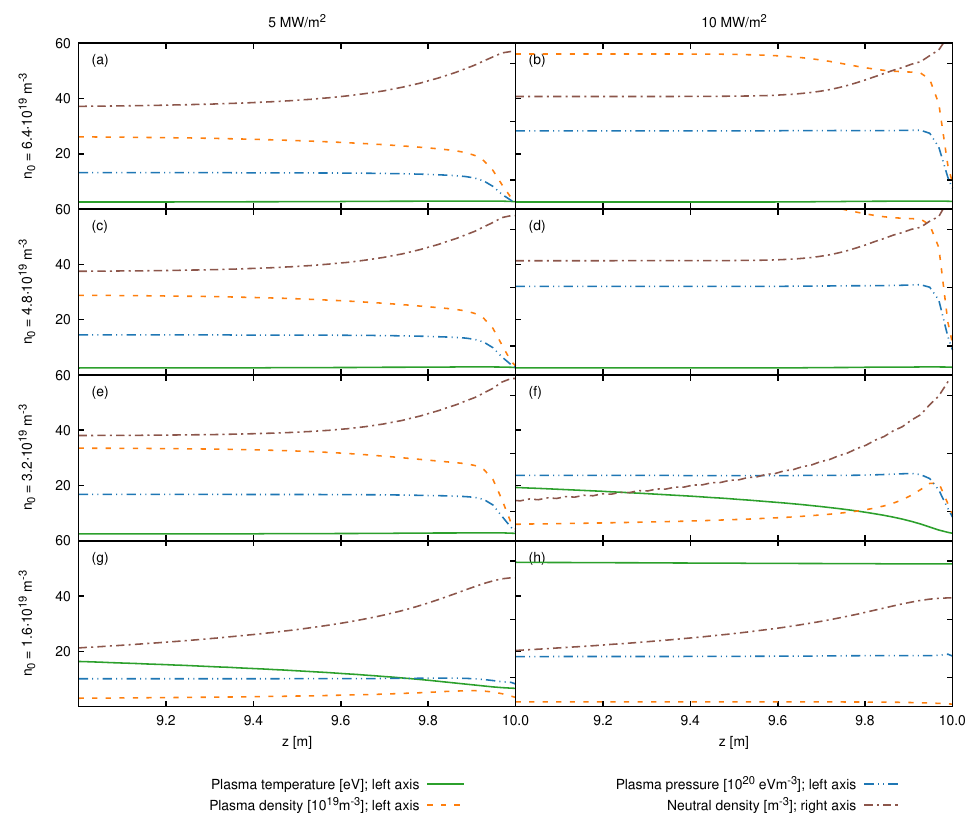}
  \caption{
    Zoom-in of the last m in front of the target of
    fig.~\ref{f:bg}. See fig.~\ref{f:bg} for details.
    }\label{f:bg2}
\end{figure}
the 1D simulation were run
to steady state. Fig.~\ref{f:bg} shows some of the obtained backgrounds.
With the exception of the bottom right figure, a significant
temperature drop towards the target is observed.
The plasma pressure, plotted with a blue dashed double dotted line
stays mostly constant, until shortly before the target.
The target pressure in fig~\ref{f:bg2} (a,c) drops to 20\,\% of the
upstream value, while in  (g) the pressure stays basically
constant along the flux tube, while
in (b) the pressure drops to around $\frac 1 3$ of the upstream value.
As the upstream plasma density increases, the region of increased neutral
density extends along the field line, where the plasma temperature
is low, and the plasma
density is increased. As long as the pressure stays constant, a
decrease in plasma temperature coincides with an increase in plasma
density to conserve pressure. $0.5$\,m in front of the target the neutral density
increases further, which causes the  plasma pressure to decrease
several cm
in front of the target.
\begin{figure}
  \centering
  \includegraphics[width=.7\linewidth,page=17]{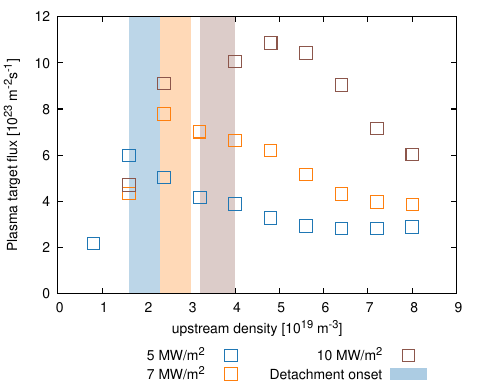}
  \caption{Particle target flux roll-over plot for different energy influxes,
    obtained with the described 1D model.
    Some of the profiles are shown in fig.~\ref{f:bg}.
    The shaded area represent detachment onset, according to the
    onset of pressure drop.
  }\label{f:roll}
\end{figure}
The dependence of the particle target flux on the upstream density is shown in
fig.~\ref{f:roll}. For the 7 and 10\,MW/m$^2$ cases the particle target
flux increases initially before it decreases for higher densities. In
the 5\,MW/m$^2$ case the roll-over happens below
24$\cdot10^{18}$\,m$^{-3}$.
This means that most of the 5\,MW/m$^2$ cases are detached, while only
the higher density cases at 10\,MW/m$^2$ are detached. Detachment
on-set, featuring pressure drop, is indicated by the shaded areas for
the respective energy influxes.

While this code produces detached solutions, the details may vary
compared to more complex codes, such as SOLPS, due to the
additional neutral physics, as for example no molecules are included,
or the lack of the 2nd or 3rd dimension~\cite{wischmeier09a}.
Further kinetic neutral effects, molecular effects, as well as effects
due to a full geometry are neglected.
As the three main interaction between plasma and neutrals are
included, namely charge exchange, ionisation and recombination,
It is expected that this model still captures the general trend and as
it is the first study of interaction between filaments and detachment
provides motivation for further, more complete studies.
Thus the results are not expected to accurately predict experimental
observation, but capture the lowest order effect of detachment on
filament dynamics and thus give qualitative useful results concerning
the phenomenology of filament motion in detached conditions.

\section{Filament evolution}\label{s:evo}
Filaments were seeded on the backgrounds discussed in sec.~\ref{s:bg}.
They
were seeded as a Gaussian perturbation in the drift plane, and with a
tanh shape in the parallel direction, where the length was chosen as
5\,m. In the perpendicular direction a Gaussian width of 2\,cm was
chosen for the initial studies~\cite{schworer18a}, which was later
varied to study the role of $\delta_\perp$.

The perturbation for density and temperature was chosen equal to the
upstream background value $n_0$ and $T_0$. This keeps the relative
perturbation $\frac{\delta_n}{n_0}$ as well as $\frac{\delta_T}{T_0}$
in all cases upstream constant, at
$\frac{\delta_n}{n_0}=\frac{\delta_T}{T_0}=1$.

The filament was evolved and the centre of density was tracked in the
drift plane by taking the centre of mass above a threshold. This is
described in more detail in our previous paper~\cite{schworer18a}.

\subsection{Direct influence of neutrals}\label{s:di}
\begin{figure}
  \centering
  \includegraphics[width=\linewidth,page=2]{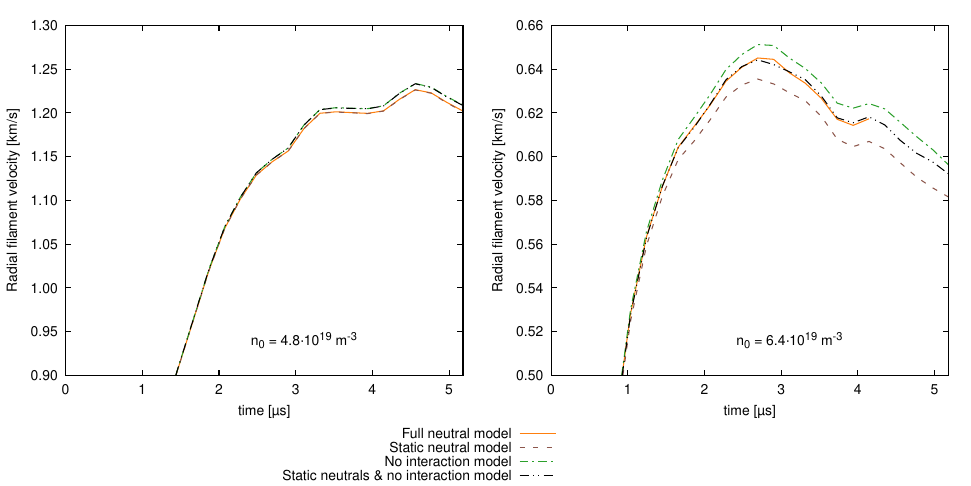}
  \caption{Time evolution of the radial velocity of a filament seeded
    on the background with 10\,MW/m$^2$ and
    $n_0=48\cdot10^{18}$\,m$^{-3}$ (left) as well as
    $n_0=64\cdot10^{18}$\,m$^{-3}$ (right), shown in
    fig.~\ref{f:bg}(d,b).  Results for different neutrals model are
    shown, see legend. The initial accelleration is not shown to make
    the differences between the models more visible.
  }\label{f:dir1}
\end{figure}
Fig.~\ref{f:dir1} shows the time evolution of the radial velocity of a
filament seeded on the high density backgrounds \dens{48} and
\dens{64} in the \MW{10} case, shown in fig.~\ref{f:bg}(b,d).  The
no-interaction models in all cases show higher radial filament
velocity then the version that includes the vorticity-neutral
interaction, however the effect is below 1\,\% in the \dens{64} case
and not noticeable in the lower density cases.

\begin{figure}
  \centering
  \includegraphics[width=\linewidth,page=1]{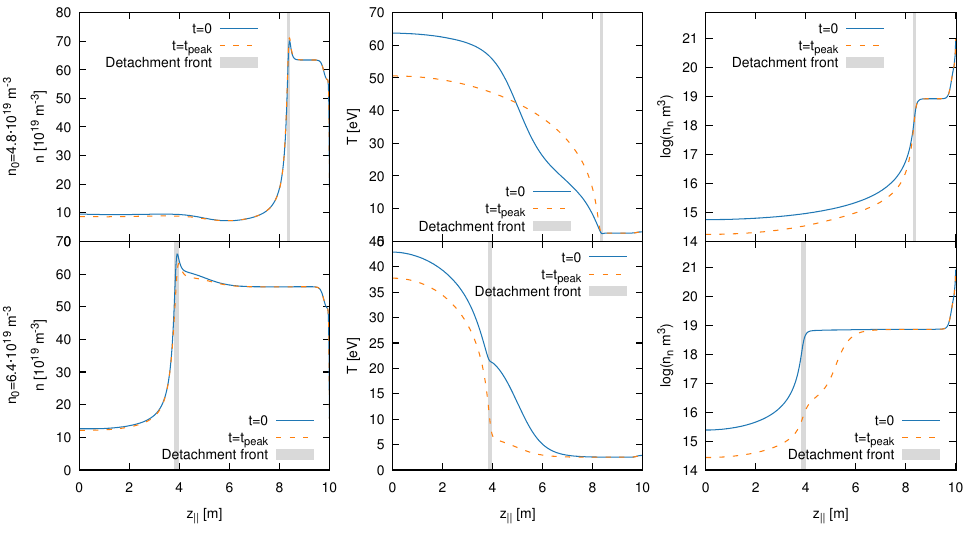}
  \caption{Plasma profiles at the centre of the filament. Shown in
    blue are the profiles just at the beginning of the simulation,
    before the neutrals could adjust to the perturbation. Shown in red
    are the profiles after $\approx$ 4\,µs.
    Shown is the plasma density (left), electron temperature (centre)
    and neutral density (right). On the top is the \dens{48} case, and
    on the bottom the \dens{64} case, both with 10\,MW/m$^2$ energy
    influx, as shown in fig.~\ref{f:bg}(d,b)
    The detachment front is shaded in grey.
  }\label{f:dir1exp}
\end{figure}

In the \dens{48} case, keeping the neutrals static decreases the
velocity by less then 0.1\,\%. In the \dens{64} case where the cold
detachment front is within the seeded filament, the
static neutral simulation over-estimates the influence of the
neutrals, resulting in an decreased radial filament velocity by
about 1.5\,\%.
This can be explained by looking at the plasma profiles,
shown in fig.~\ref{f:dir1exp}.
In the \dens{48} case the cold plasma region does not extend into the
seeded filament, and thus the neutral density is rather low in the
seeded filament. In the \dens{64} case, the filament is seeded within
the cold plasma region which includes a high neutral density.  Thus in
the \dens{64} case a significant amount of neutrals are ionized by the
filament, unlike in the \dens{48} case.



\subsection{Dependence on detachment}\label{s:de}
Compared to the previous study, in attached conditions, the
velocity of filaments in these higher density simulations is
decreased~\cite{schworer18a}.
Previously the slowest filament velocity reached was around
550\,m/s at an target temperature of 0.8\,eV.
Higher target temperatures resulted in faster filament velocities.
In the simulations presented here the target
temperature does not drop below 2.5\,eV, as the recombination acts as a
heat source for the electrons at low temperatures.
\begin{figure}
  \centering
  \includegraphics[width=.7\linewidth,page=6]{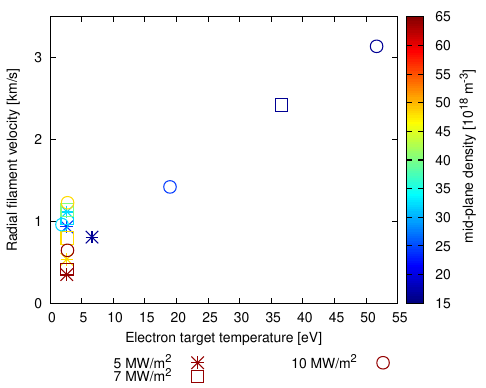}
  \caption{Peak radial filament velocity for the backgrounds shown in
    fig.~\ref{f:bg} versus target temperature of the background. The
    target temperature dependency breaks down for target temperatures
    below 5\,eV.
  }\label{f:vstt}
\end{figure}
Fig.~\ref{f:vstt} shows the peak radial filament velocity versus the
target temperature. While for the high target temperature cases our
previous findings of a target temperature dependence are reproduced~\cite{schworer18a},
this does not hold for target temperatures below
$\sim$5\,eV.
The main differences to the previous study is the increased plasma
density, as well as the Franck-Condon energy source from ionisation.

\begin{figure}
  \centering
  \includegraphics[width=.7\linewidth,page=10]{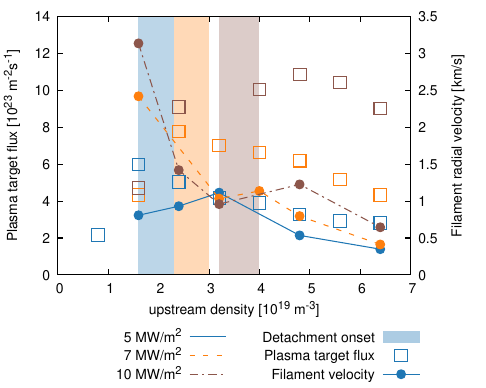}
  \caption{On the left axis is the
    particle target flux of the associated background profiles.
    On the right axis is the peak radial filament velocity for
    different plasma backgrounds. The lines are to guide the eye, and
    do not imply that the transition between the points would be
    linear. The shaded are represent detachment onset for the
    different energy influxes. 
    A general decreasing trend of the radial velocity with increasing
    density is observed, except at the flux roll-over point, where the
    filaments become temporarily faster.
  }\label{f:roll-full}
\end{figure}
\begin{figure}
  \centering
  \includegraphics[width=.7\linewidth,page=2]{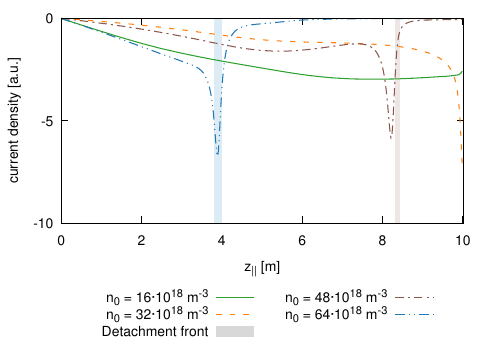}
  \caption{Current density at the centre of the filament after
    $\sim$4\,µs for the 10\,MW/m$^2$ cases shown in
    fig.~\ref{f:bg}(b,d,f,h).
    In the detached cases, currents near the target are strongly
    suppressed.
    The detachment front is shaded in the respective colour.
  }\label{f:nocur}
\end{figure}

To understand the dependence of the filament velocity in detached
conditions, fig.~\ref{f:roll-full} shows the filament velocity as a
function of upstream density, as well as particle target flux of the associated
backgrounds.
The filament velocity generally decreases with increasing density.
The 5\,MW/m$^2$ case has an exception for densities less then \dens{32}, the 7\,MW/m$^2$
case between \dens{32} and \dens{40} and the 10\,MW/m$^2$ case between
\dens{32} and \dens{48}.
The particle target flux roll-over is at \dens{16}, \dens{24} and \dens{40}
respectively.
In all three cases, with increasing density the target flux roll-over
and pressure drop happen, and at even higher densities the radial
velocity increases temporarily.
This suggests that after detachment is reached, the filaments get
faster, before the velocity begins to decrease again.
At this point, a cold and dense plasma in front of the
target is building up and with increased density
as well as decreased temperature, the resistivity is significantly
increased, which suppresses currents in the filament reaching the
target. Fig.~\ref{f:nocur} shows the current density in the centre of
the filament, for the 10\,MW/m$^2$ cases.
While in the attached cases, the currents flow to the target on the
right hand side, in the detached cases the currents are only flowing
to the edge of the cold plasma region, where strong currents due to
the large density gradients are observed.
The increase of the filament radial velocity with increased
resistivity was predicted~\cite{krasheninnikov08a,easy16a,garcia09a}. The decrease of
the filament velocity
with increasing density is at least partially due to a decreased drive
as the temperature is decreased.
In addition to an decreased drive, the plasma viscosity increases,
which also reduces the filaments velocity.
Thus in general the decreased drive and the increased viscosity are
dominating over the increased parallel resistivity for most densities,
with the exception after detachment is reached, where the velocities
temporarily increase and the increased resistivity dominates over
the other changes.


\subsection{Dependence of critical size}\label{s:ds}
The sink for the vorticity drive, and thus the dynamics of filaments
strongly depends on the size of the filament with respect to the
critical size $\delta^*$. For small filaments, i.e. $\delta_\perp <
\delta^*$, the vorticity is balanced mostly in the drift plane, and the
filament velocity increases monotonically with size. For large filaments,
the vorticity is balanced predominantly via parallel dynamics, and a
monotonic decrease with filament size is expected.

\begin{figure}
  \centering
  \includegraphics[width=.7\linewidth,page=2]{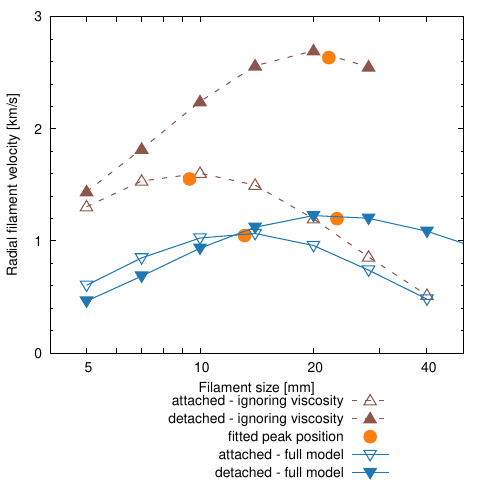}
  \caption{Peak radial filament velocity for different sized
    filaments.
    The blue, filled symbols are in detached conditions (\MW{10},
    \dens{32}, fig.~\ref{f:bg}(f)), while the
    red, open symbols are in attached conditions (\MW{10}, \dens{48},
    fig.~\ref{f:bg}(d)).
    The downward pointing triangles are from simulations using the full
    model, whereas the upward pointing triangles are from simulations
    ignoring the perpendicular viscosity.
    The fitted position of the critical size is shown as red dots.
  }\label{f:peakboth}
\end{figure}

To further understand the influence of detachment on filament
dynamics, different sized filaments where seeded on the background
profiles.
The result of the filament size scan is shown in fig.~\ref{f:peakboth}.
Also shown is the critical size, as determined by fitting a quadratic
function in log-space using gnuplot~\cite{gnuplot503}.
Additionally to the full model, a second case is run where the
perpendicular viscosity is set to zero.
While the perpendicular diffusion constants are physically motivated,
the real value is not known. Some models don't include viscosity or
diffusion at all~\cite{angus12a,lee15a}.
The previous study found a strong viscosity dependence of the filament
dynamics, thus the viscosity dependence was once more investigated.
The neo-classical viscosity values were assumed to be the
upper bound and setting the viscosity to zero is a natural lower bound.
Thus the resulting error of this uncertainty is expected to be bound
by the full model on one side, and by the no-viscosity case on the
other hand.

In the filament size scan in attached conditions, the regime
transitions from the viscous regime for small filaments to the sheath
limited regime for large filaments, where sheath currents are the main
current closing mechanism. In detached conditions, for small
filaments, the vorticity drive is mainly balanced by viscosity. For large
filaments, sheath current can no further close the vorticity drive, as
sheath currents are strongly suppressed, as shown in
fig.~\ref{f:nocur}.
As mentioned earlier, the filament scaling expected for neutrals does
not have a size dependence.
As with increasing size the monotonic decrease is
retained, while the high resistivity strongly reduces sheath currents,
sheath currents are not fully suppressed.
\begin{figure}
  \centering
  \includegraphics[width=\linewidth]{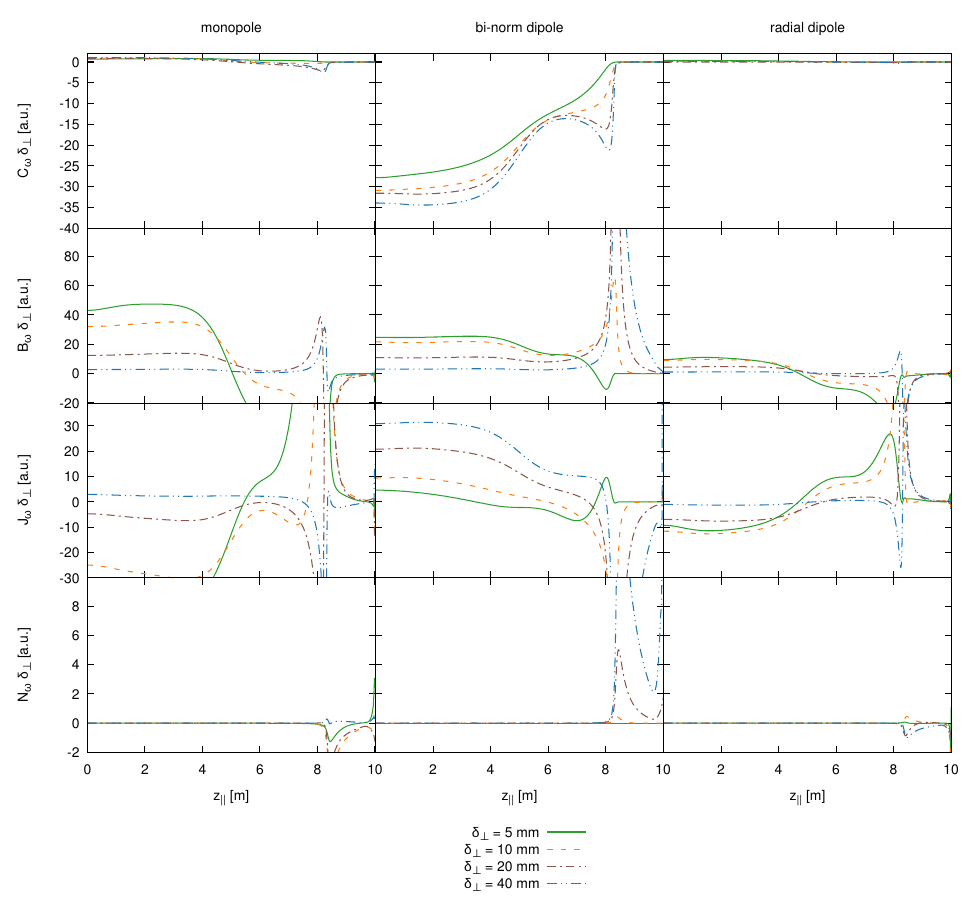}
  \caption{Contributions of the vorticity equation for the
    \dens{48} \MW{10} case for different sized filaments. The
    contributions are split in the base-functions discussed in
    sec.~\ref{s:model}. The advective contribution isn't shown, as
    it's magnitude is well below 3\,\% of the other contributions.
  }\label{f:vc1}
\end{figure}
Looking at the contributions in the vorticity equation, it can be
observed that the viscosity contribution is increasing for filaments
until $\delta_\perp\approx20$\,mm. While the viscosity contribution
within the filament decreases with increasing size, the contribution
at the detachment front shows a more complicated behaviour, as shown
in fig.~\ref{f:vc1}. While for
small filaments the vorticity is balanced locally, for large ones the
vorticity drive is advected towards the sheath, and as the viscosity is more
effective at higher density and lower temperature, the viscosity still
provides a significant closing mechanism at the front of the detached
volume, but decreases quickly with increasing size due to the
$\delta_\perp^{-2}$ dependence.

As the vorticity sink provided by viscosity is decreasing with increasing size, the
impact of the neutrals is increasing. For the $\delta_\perp=20$\,mm
filament, the neutrals contribution is roughly $\frac{1}{60}$ of the
viscosity contribution, while in the $\delta_\perp=40$\,mm case the
the neutrals have nearly doubled their impact, and the neutrals
contribution is about $\frac{1}{20}$ of the viscosity contribution.

In the full model, the filaments of 14\,mm and larger are faster in
the detached condition than in attached conditions.
This is expected, as small filaments are not influenced by sheath
currents, and the increased viscosity, $\mu_\omega\propto n
T^{-\frac 1 2}$ as mentioned in eq.~\eqref{eq:mu}, resulting from
the higher density
in detached state will thus reduce the velocity the filament
achieves. For large
filaments, the suppression of sheath currents is much more important and the
effect of viscosity is weaker, thus they are faster in detached conditions.
In the case without viscosity, the filaments of all sizes, down to
5\,mm are faster in the detached condition than in the attached case.
This indicates that even for 5\,mm filaments, in the absence of
perpendicular viscosity, parallel dynamics plays
a role, as a significant part of the vorticity drive is advected towards the
target. In the low resistivity, attached case, sheath currents provide a monopole
contribution, and the $E\times B$ term causes a significant
contribution to the vorticity sink.  In the high resistivity case, the
$E\times B$ contribution is significantly reduced. In that case
parallel advection and diffusion are preventing further acceleration
of the filament.
Therefore in the absence of viscosity, even the small filaments are
influenced by the sheath conditions, thus filaments of all sizes
achieve higher velocities in detached conditions, where sheath effects
are suppressed.

The full model as well as the no viscosity model shows an increase
in velocity with detachment for larger filaments.
Further, the critical size $\delta^*$ increases significantly with
detachment. Note that $\delta^*$ is defined as the perpendicular
size where filaments are fastest.

\begin{figure}
  \centering
  \includegraphics[width=.7\linewidth,page=1]{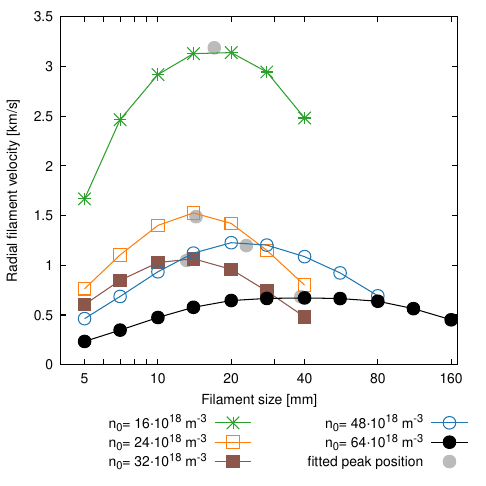}
  \caption{Peak radial filament velocity for different sized
    filaments for different backgrounds.
    The backgrounds have an energy influx of \MW{10}, and the
    densities range from \dens{16} to \dens{64}.
  }\label{f:peakfull}
\end{figure}

To put these results further into context, the scan has been extended
to more densities, as shown in fig.~\ref{f:peakfull}
\begin{figure}
  \centering
  \includegraphics[width=.7\linewidth,page=1]{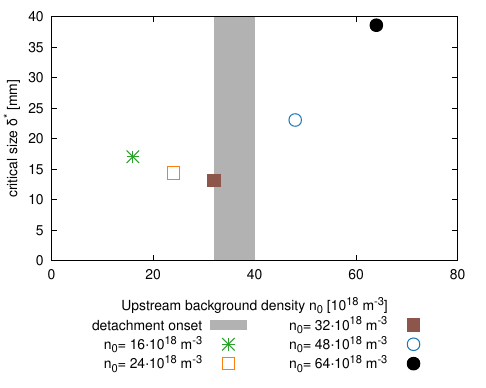}
  \caption{Critical size of the different densities shown in
    fig.~\ref{f:peakfull}.
    The shaded region denotes the onset of detachment.
    The backgrounds have an energy influx of \MW{10}, and the
    densities range from \dens{16} to \dens{64}.
  }\label{f:peakfull2}
\end{figure}
The density dependence of filaments of $\delta_\perp=20$\,mm has
already been discussed in section~\ref{s:de}.
The trend is the same for larger filaments as well, while small filaments
(10\,mm and smaller) show a monotonic decrease with increasing
density.
In attached conditions the critical size $\delta^*$ decreases with
increasing density. This is in contrast to our previous study, where a
increase with increasing density was observed. The main difference
between the two studies is, that here the energy influx was kept
constant, whereas in the previous study, the upstream temperature was
kept constant~\cite{schworer18a}. In the case of a fixed heat flux, an
increased density results in an decreased temperature, due to pressure
conservation.
It seems thus that the decreasing
temperature reduces the critical sizes stronger, then the density
increases the critical size.
As detachment is reached, the critical size $\delta^*$ shifts
dramatically to
larger sizes. This can be explained by the suppression of sheath
currents, which allows larger filaments to reach higher
velocities. As larger filaments are faster, the critical size is
shifted towards larger sizes.
Fig.~\ref{f:peakfull2} shows the plot of the critical size
plotted against the respective density. The critical size was
determined by fitting a quadratic function against the radial velocity
versus logarithm of the perpendicular size using
gnuplot~\cite{gnuplot503}.
In the \dens{64} case the critical size is further increased with
respect to the \dens{48} case, which is already detached. This
demonstrates the importance of how far the detachment has moved
upstream, which roughly linearly increases the collisionality
integrated along the flux tube.

Based on these results, it is expected that for experimentally observed filaments
in detached conditions are generally slower than in fully attached
conditions, however not a simple relationship with detachment is
expected, but rather filament size $\delta_\perp$ with respect to the
critical size $\delta^*$ needs to be taken into account.

\subsection{Rigidness of filaments}\label{s:r}
Filaments have been observed to bend in electro-magnetic models, where
the pressure perturbation is not any more much smaller than the magnetic
pressure~\cite{lee15a}.
The model used here however uses the electrostatic assumption, and in
previous studies the filaments have been observed to be rigid.
\begin{figure}
  \centering
  \includegraphics[width=\linewidth,page=1]{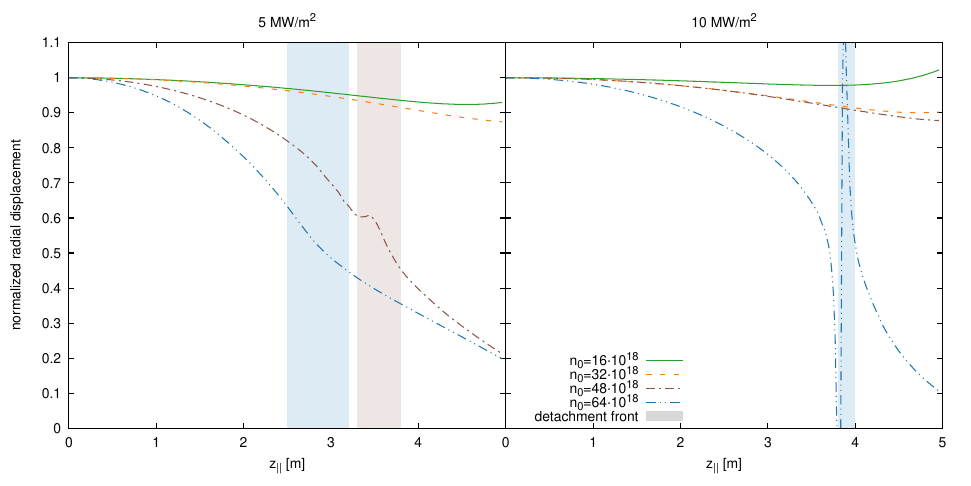}
  \caption{Dependence of the filaments radial displacement after
    4\,µs on the parallel
    direction $z_\parallel$ for the different backgrounds in
    fig.~\ref{f:bg}.
    The displacement was normalized to the displacement at
    $z_\parallel=0$. Note that only the upper half of the domain is
    plotted, where the filament was seeded.
    The detachment front is represented by a shaded area.
    In the \dens{64} \MW{10} case the detection did not work reliably
    around $z_\parallel \approx 4$\,m, as the background density
    features strong parallel gradients, thus small movements of the
    detachment front making the detection of the centre of the
    filament unreliable.
  } \label{f:ydep}
\end{figure}
The radial displacement of the filaments seeded on the backgrounds shown
in fig.~\ref{f:bg} has been computed for the different $x$-$y$-slices
along $z_\parallel$ and is shown in fig.~\ref{f:ydep}.
In attached conditions, the velocity has no $z_\parallel$ dependence,
and the filament moves rigidly.
Also in detached conditions, the filaments in \dens{48} \MW{10} as
well as \dens{32} \MW{5} still move
rigidly.
Only the filaments in \dens{64} with \MW{10} and \MW{5} as well as
\dens{48} \MW{5} show bending.
In addition all showing a low radial velocity, the filaments that bend
have in
common that they were seeded partially within the detached
region. This prevents currents within the filament to flow freely,
which explains why the vorticity within the filament is strongly
dependent on the position within the filament.

In fig.~\ref{f:ydep} the filaments that are in a detached region
show significant bending. Around the detachment front, the filament
has only move 60\,\% of the upstream value. Within the detached
region, the retardation is even more prominent. The bended filaments
are seeded on the backgrounds shown in fig.~\ref{f:bg}(a,b,c).  Whilst
this is an idealised scenario as the filaments are seeded partially
within the detached region, the deviation of filaments from field
alignment in detached plasmas may be a useful result for
experimental comparison.

\section{Discussion}\label{s:dis}
Similar to the previous study, strictly in attached conditions,
keeping the neutrals fixed at their
background values has only a minor impact on the dynamics of
filaments.
If filaments are able to penetrate into a detached
region, they can ionise significant amount of neutrals, which may be
relevant to detachment burn-through studies, however.
Burn-through is not studied in this paper, as the neutral code was
unstable close to re-attachment.

Easy~\etal predicted that a cold divertor could increase the radial
velocity of filaments, as well as the critical size $\delta^*$ due to
a rise in collisional resistivity~\cite{easy16a}. While both an
increase in radial velocity
as well as an increase of $\delta^*$ has been observed as detachment
is reached, the increase of the radial velocity is rather weak, and a
hotter divertor yields higher radial velocities in general.
It has been suggested that detachment, or more generally a high
collisionality in the divertor could be the cause of SOL flattening, as
the high resistivity could prevent sheath currents, and thus
increase the radial filamentary transport~\cite{carralero17a,kuang19a}.
This study shows that collisionality may be responsible for a
change of filament dynamics, rather then e.g. the increased neutral
density.
However, the increase of filament velocities with
detachment is well below the radial velocities in sheath limited
conditions, 
so this study does not support the hypothesis that shoulder formation
is the result of high divertor collisionality and increased filament
transport fluxes. The increase with detachment was significantly
larger then the errors due to solver accuracy, finite discretisation
or finite domain size effects, which were found to be below 2\,\%.


Furthermore, as a high plasma density as well as a low plasma
temperature is required to achieve a sufficiently high resistivity to
sufficiently prevent sheath currents to influence the filament dynamics, this will most likely
not be the case in the far SOL, thus even if the acceleration caused by
detachment is higher then suggested by these simulations, the velocity
should quickly drop, once the filament is connected to a lower density
region and thus can connect to the target.



After detachment occurs, a strong increase of the critical filament
size $\delta^*$
with increasing density is
observed. In attached conditions a weak reduction of the critical size
with increased density is observed, which is associated with a
reduction of the temperature
as the energy influx is kept constant.
After detachment occurs, cold and dense plasma at the target is
suppressing sheath currents, and with increasing density the
detachment volume is increasing, thus shifting the critical size
to larger values. While
neutrals can partially compensate for the lack of sheath currents, the
vorticity sink due to neutrals is over an order of magnitude smaller
than the sink due to viscosity.

The observed filament bending agrees with Easy~\etal\cite{easy16a}
where bending was observed if the
resistivity was uniformly enhanced, whereas a localized resistivity at
the target resulted in a rigid filament~\cite{easy16a}.
Note that this bending mechanism would only be expected in
experiments, if the filament is within the cold, dense plasma
region. The bending was only observed in cases where the
detachment front has moved quite far upstream, and the filament was
within the detached plasma.

Easy~\etal\cite{easy16a} increased the resistivity artificially by up
to a factor of 10\,000
and estimated that the temperature at the divertor would need to drop
to 0.086\,eV compared to the 40\,eV reference case~\cite{easy16a}. The
estimates assumes that the resistivity has a temperature dependency
$\nu_\parallel \propto T^{-3/2}$. However, the resistivity also has a
density dependence i.e. $\nu_\parallel \propto n \cdot T^{-3/2}$.  If
we assume that the pressure $n\cdot T$ is preserved (same upstream
conditions but colder divertor) the scaling can be expressed again in
terms of the temperature: $\nu_\parallel \propto T^{-5/2}$.  Thus
achieving the 10 000 increase only requires 1\,eV - which is much
closer to the temperatures reached here and may explain why the
bending effect is seen in the detached simulations.

While Easy~\etal\cite{easy16a} predicted that the velocity will be increased with
increased resistivity, the self-consistent simulations conducted here
show
that the increase with detachment is significantly smaller, than the
net decrease with respect to attached simulations, where target temperatures above
25\,eV reach velocities exceeding 2\,km/s - whereas in detached
conditions only about 1\,km/s was reached.
However, a change in the transition of filament dynamics from inertial
or viscous to sheath-limited is observed and the critical size of
filaments $\delta^*$ increases dramatically after detachment.
This is in agreement with experimental observation, that an
increased collisionality does not result in an increase of the
radial filament velocity~\cite{vianello19a}.
Rather an increase of the filament size coincides with shoulder
formation~\cite{vianello19a}. 

\section{Summary}\label{s:sum}
The paper presents attached as well as detached 1D parallel plasma
profiles and the dynamics of filaments in a 3D slab geometry seeded on
these profiles.
The detachment features particle target flux roll-over as well as a significant
plasma pressure drop.
Detachment was achieved by using a refined neutral model, which is
able to capture and evolve steep gradients in the neutral density,
which is observed near the target.


In terms of dependence of the filament dynamics on the background
conditions, a general decreasing trend of radial filament velocity
with increasing plasma density is observed. This trend is temporarily
broken as detachment is reached, where especially filaments larger
then $\delta^*$ are faster then before detachment.
This is caused by the higher parallel resistivity.
This also causes an increase of the critical size, which further
increases as the detachment front moves further upstream an the
integrated resistivity increases.

While detachment can increase the radial velocity, the observed radial
filament velocities in hotter, attached plasmas are still faster then
the ones observed in detachment.

\nocite{tange18a}
\section{Acknowledgement}
This work has been carried out within the framework of the
EUROfusion Consortium and has received funding from the
Euratom research and training programme 2014-2018 and
2019-2020 under grant agreement No 633053. The views and
opinions expressed herein do not necessarily reflect those
of the European Commission.
This work used the
EUROfusion High Performance Computer
(Marconi-Fusion) through EUROfusion.

\bibliographystyle{iaea_misc}
\bibliography{phd}

\end{document}